\documentclass
[preprint,10pt,twocolumn,prl,letterpaper,noshowpacs,balancelastpage]{revtex4}%
\usepackage{amsfonts}
\usepackage{amsmath}
\usepackage{amssymb}
\usepackage[dvips]{graphicx}
\usepackage{hyphenat}%
\providecommand{\U}[1]{\protect\rule{.1in}{.1in}}

\begin{document}
\preprint{ }
\title{Onset of nonequilibrium in a driven Anderson insulator}
\author{Z. Ovadyahu}
\affiliation{Racah Institute of Physics, The Hebrew University, Jerusalem 9190401, Israel }


\begin{abstract}
The onset of nonequilibrium in a driven Anderson-insulator is identified by
monitoring the system with two-thermometers. Features of nonequilibrium appear
at surprisingly weak drive intensity demonstrating, among other things, that
conductivity may not be a reliable thermometer for ensuring linear-response
conditions. In addition, the spectral contents of the applied field could be
more important to take the system out of equilibrium than its absorbed power.
Ensuing hot-electron transport effects and the nontrivial role phonons play in
driven quantum systems are pointed out.

\end{abstract}
\maketitle

\textit{Below 1K, you know the temperature if you have one thermometer; when
you have two, you don't}. This adage is recalled whenever a sample resistance
fails to follow the expected dependence on temperature given by the
fridge--installed thermometer. Such an event is often related to an influx of
energy causing a lack of detailed-balance with the bath. The sample may still
be in a steady-state but not in thermal-equilibrium. Considerable attention
has been given to nonequilibrium states, created by a variety of means, to
look for novel phenomena \cite{1,2,3,4,5,6}. A non-equilibrium steady-state
(NESS) is a common occurrence in low temperature transport experiments making
it a natural testbed for these studies.

Monte-Carlo simulations of driven disordered systems supported the intuitive
notion that the electrons may be assigned an effective temperature
T$_{\text{eff}}$ distinctly different from the phonon temperature \cite{7,8}.
The effective temperature of the electrons was defined by fitting their energy
distribution to the Fermi-Dirac expression \cite{7,8}. The relevance of an
effective temperature under drive by external fields is one of the issues
tested experimentally here.

In this note, we outline a method to detect the onset of nonequilibrium in a
driven electronic system by comparing the readings of two thermometers. These
are attached to the electronic conductance of an interacting
Anderson-insulator, aka the electron glass \cite{9,10,11,12,13,14,15,16,17}.
Once driven by external fields, both thermometers read higher temperatures
than that of the bath. As the drive power increases from the lowest level of
the experiment, they exhibit different T$_{\text{eff}}$'s signalling a
breakaway from thermal equilibrium. The difference between the two
thermometers readings, $\Delta$T$_{\text{eff}}$ increases monotonously with
the drive intensity. Significantly, $\Delta$T$_{\text{eff}}$ turns out to be
sensitive to the spectral contents of the energy supplied by the drive, not
just its magnitude. The results shed further light on the glassy nature of the
Anderson insulating phase, and in particular, on the reason for its sluggish
dynamics vis-a-vis the much faster transitions involved in the conductance.

Samples used in this study were 20nm\ thick films of amorphous indium oxide
(In$_{\text{x}}$O). These were made by e-gun evaporation of 99.999\% pure
In$_{\text{2}}$O$_{\text{3}}$ onto room-temperature Si wafers in a partial
pressure of 3x10$^{\text{-4}}$mBar of O$_{\text{2}}$ and a rate of 0.3$\pm
$0.1\AA /s. The Si wafers (boron-doped with bulk resistivity $\rho\leq
$2x10$^{\text{-3}}\Omega$cm) were employed as the gate-electrode in\ the
field-effect and experiments. The samples were deposited on a SiO$_{\text{2}}$
layer (2$\mu$m thick) that was thermally-grown on these wafers and acted as
the spacer between the sample and the conducting Si:B substrate. The
carrier-concentration \textit{N} of these samples, measured by the Hall-Effect
at room-temperatures, yielded carrier concentration \textit{N}$\approx$(1$\pm
$0.4)x10$^{\text{19}}$cm$^{\text{-3}}$.There are to date six Anderson
insulators that may be used for these experiments \cite{18}. The motivation
for choosing this version of In$_{\text{x}}$O was its highly pronounced
memory-dip, which is used here as a sensitive thermometer.

Conductivity of the samples was measured using a two-terminal ac technique
employing a 1211-ITHACO current preamplifier and a PAR-124A lock-in amplifier.
Measurements were performed with the samples immersed in liquid helium at
T$\approx$4.1K held by a 100 liters storage-dewar. This allowed up to two
months measurements on a given sample while keeping it cold. These conditions
are essential for measurements where extended times of relaxation processes
are required at a constant temperature, especially when running multiple
excitation-relaxation experiments on a given sample. The gate-sample voltage
(referred to as V$_{\text{g}}$ in this work) in the field-effect measurements
was controlled by the potential difference across a 10$\mu$F capacitor charged
with a constant current. The rate of change of V$_{\text{g}}$ is determined by
the value of this current. Except when otherwise noted, the ac voltage bias
used in conductivity measurements was small enough to ensure near-ohmic
conditions. Exciting the system by infrared radiation was accomplished by a
light-emitting AlGaAs diode operating at $\approx$0.85$\pm$0.05$\mu$m. It was
placed $\approx$15mm from and facing the sample surface. The thermometers used
for the study are two distinct properties of electron glasses; the conductance
G, and its sensitivity to a change of the carrier-concentration \textit{N}.
The latter was affected by changing the voltage V$_{\text{g}}$ between the
sample and a nearby gate electrode. A typical field-effect measurement using
this MOSFET configuration is shown in Fig.\nolinebreak\nolinebreak~1.
\begin{figure}[ptb]%
\centering
\includegraphics[
height=2.4811in,
width=3.4411in
]%
{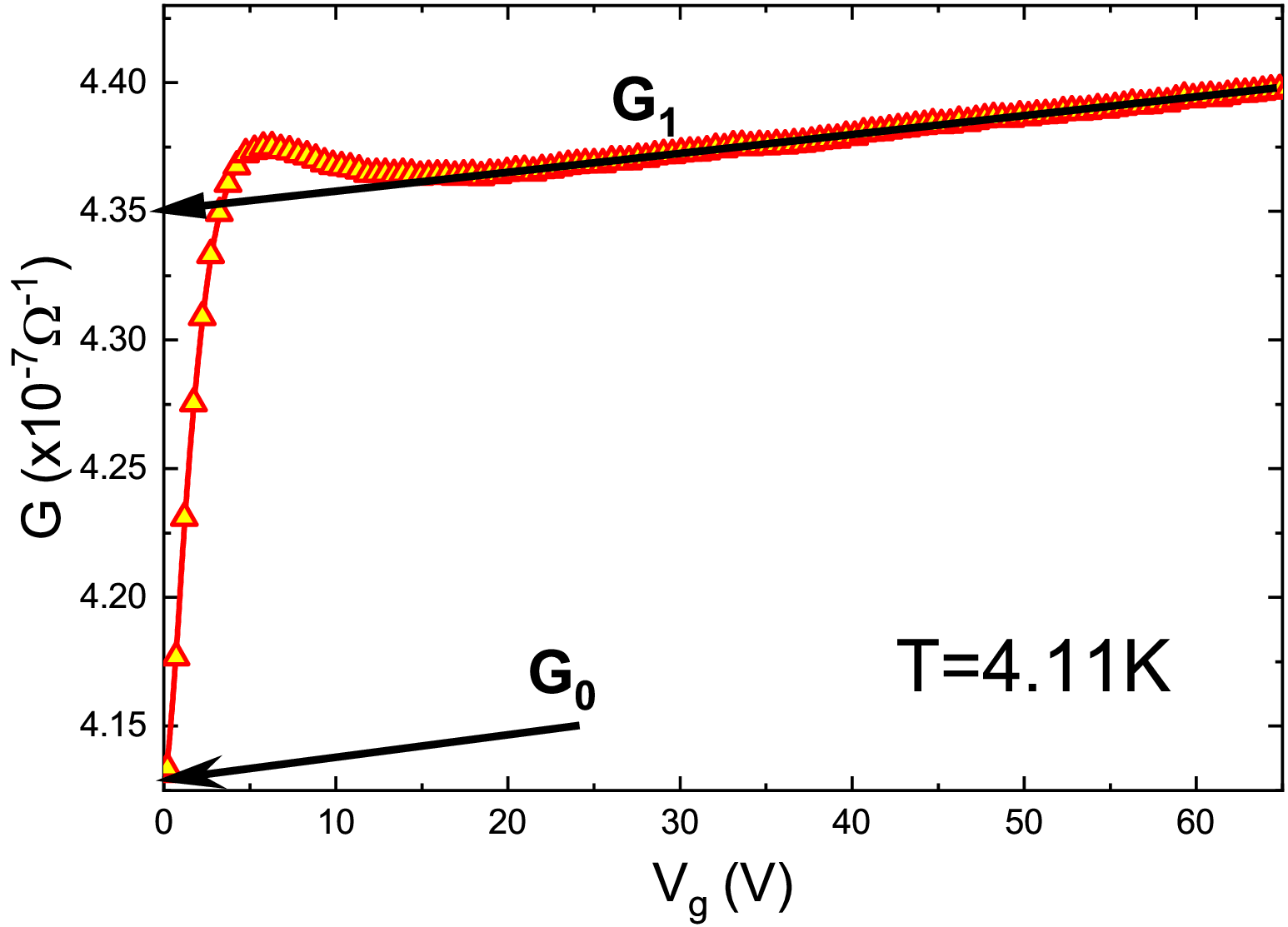}%
\caption{Fig.1- Conductance vs. gate-voltage V$_{\text{g}}$ for a 20nm thick
In$_{\text{x}}$O film with 1mmx1mm lateral size. Data were taken with a
sweep-rate $\partial$V$_{\text{g}}$/$\partial$t of 1.2V/s at a bath
temperature T=4.11K. The values of the two parameters used in the text,
G$_{\text{0}}$ and G$_{\text{1}}$, are taken as the intercepts of the arrows
with the ordinate. The "bump" at V$_{\text{g}}\approx$6V is a consequence of
charge ordering as explained in \cite{19}. }%
\end{figure}
The figure also illustrates how the parameters G$_{\text{0}}$ and
G$_{\text{1}}$, used for determining the temperature of the two thermometers
are defined. The first thermometer reading is the steady-state conductance
G$_{\text{0}}$=G(V$_{\text{g}}$=0). The second thermometer reading is $\eta
$=(G$_{\text{1}}$-G$_{\text{0}}$)/G$_{\text{0}}$ where G$_{\text{1}}$ is
determined by the construction in Fig.~1. The same sweep-rate $\partial
$V$_{\text{g}}$/$\partial$t =1.2V/s was used in all G(V$_{\text{g}}$) traces
in this work. Note that the need to complete the G(V$_{\text{g}}$) trace makes
the second thermometer `slower' than the first, a difference that plays a role
in their response. Both G$_{\text{0}}$ and $\eta$ turn out to be strongly
temperature-dependent making them sensitive `secondary' thermometers. Applying
them for temperature measurements requires calibration charts for each. These
are shown in Fig.~2 and Fig.~3 for G$_{\text{0}}$(T) and $\eta$(T) respectively.%

\begin{figure}[ptb]%
\centering
\includegraphics[
height=3.7239in,
width=3.4402in
]%
{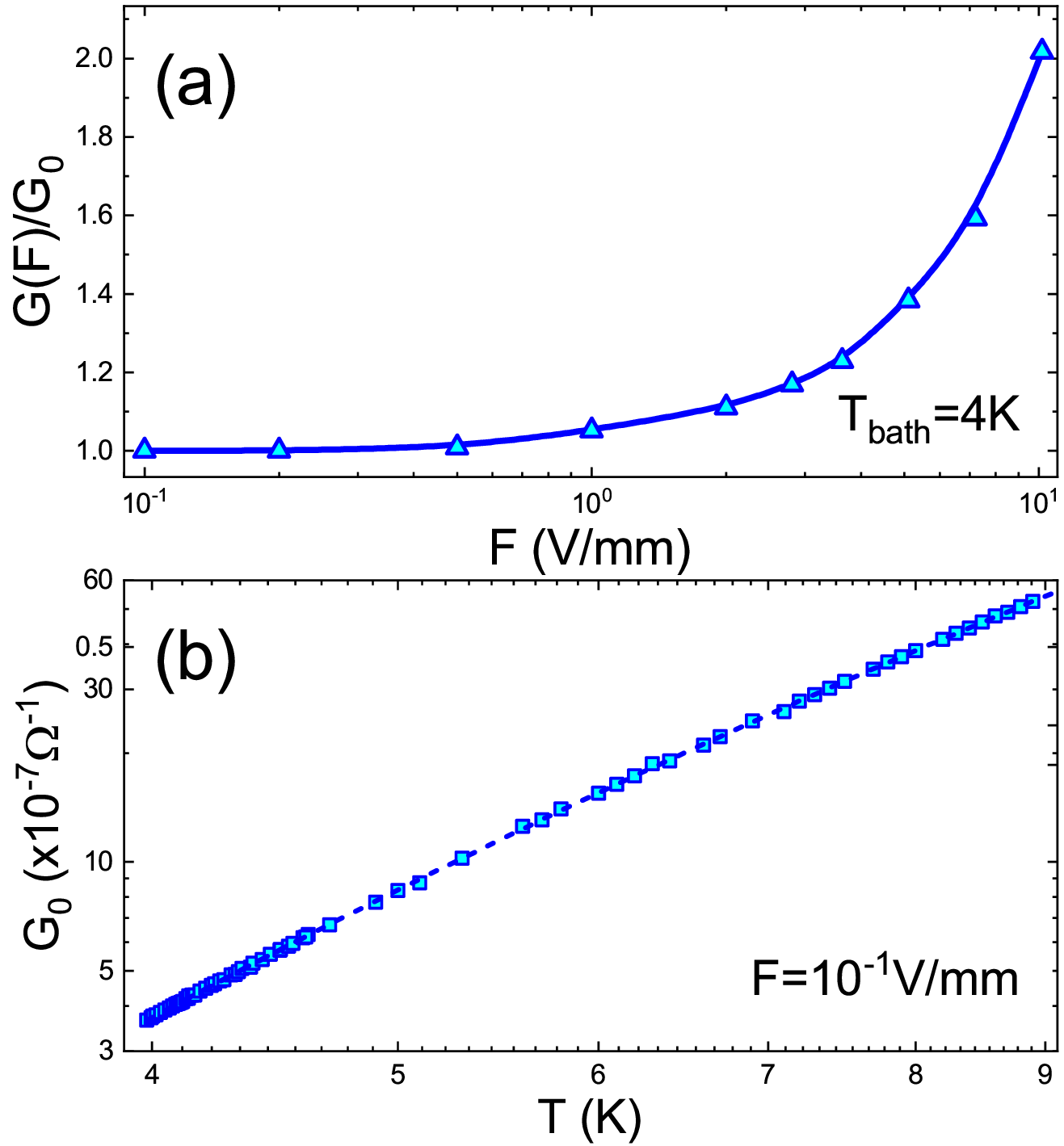}%
\caption{Pertinent transport data taken for the sample. (a): The relative
change of conductance as function of the applied electric field. (b): The
conductance vs. bath-temperature of the sample measured by two-terminal ac
technique using a bias voltage in the linear-response regime at a frequency
f=73Hz. At each point, the sample was allowed to relax for 1-2 hours to ensure
steady-state conditions. The dashed line is a fit to: G$_{\text{0}}%
$(T)$\propto\exp$[-(T$_{\text{0}}$/T)$^{\text{1/3}}$] with T$_{\text{0}}%
\simeq$5800K.}%
\end{figure}
%

\begin{figure}[ptb]%
\centering
\includegraphics[
height=2.1672in,
width=3.4411in
]%
{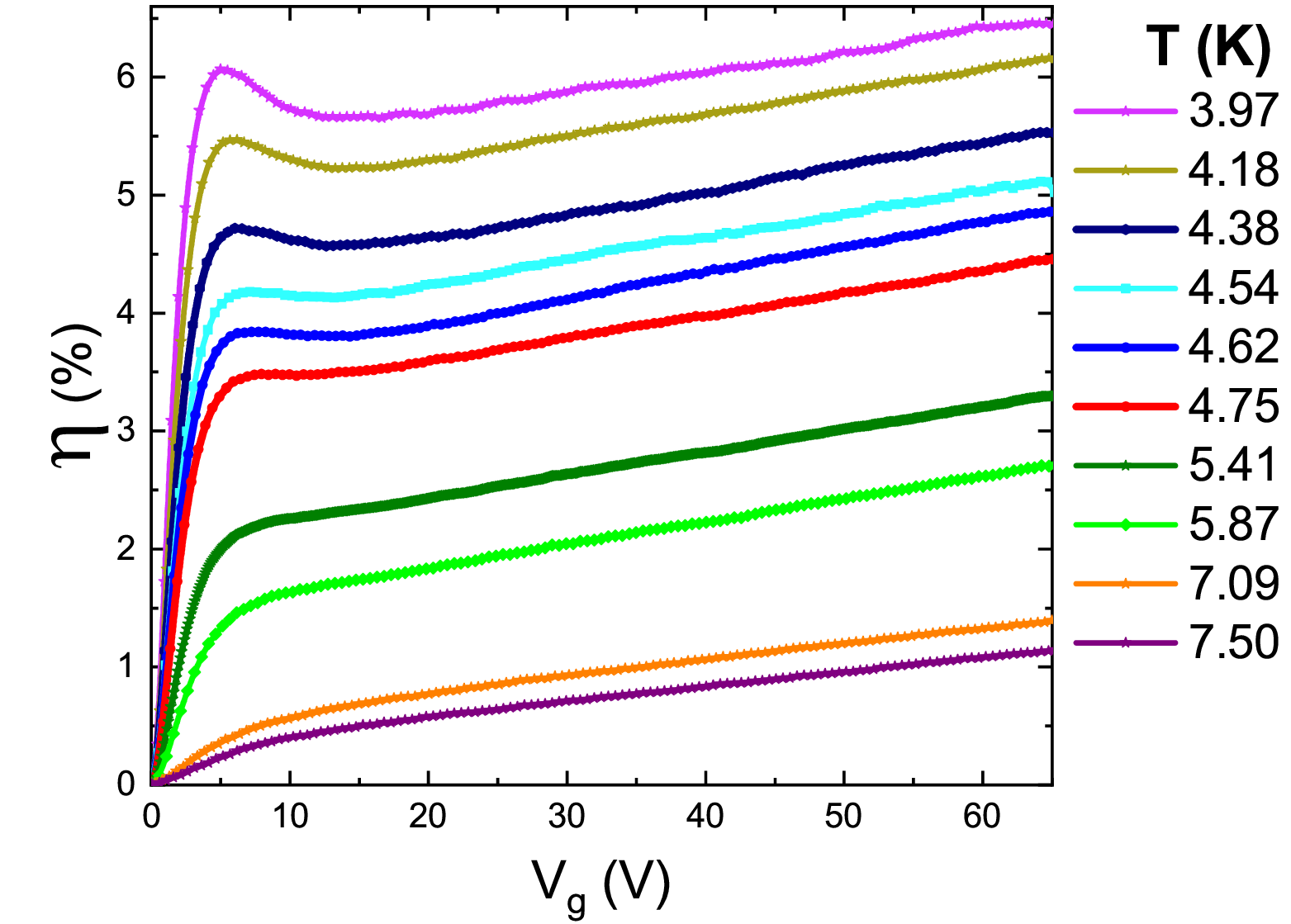}%
\caption{Conductance vs. gate-voltage G(V$_{\text{g}}$) for the In$_{\text{x}%
}$O film. Curves were taken at indicated temperatures with the same
sweep-rate: $\partial$V$_{\text{g}}$/$\partial$t=1.2V/s.}%
\end{figure}

Figure 4 shows a set of G(V$_{\text{g}}$) curves taken by using non-Ohmic
fields to measure both, the sample conductance and its associated field
effect. All other aspects of the measurements protocol were identical to those
used in obtaining the data set in Fig.~3.%

\begin{figure}[ptb]%
\centering
\includegraphics[
height=2.111in,
width=3.4411in
]%
{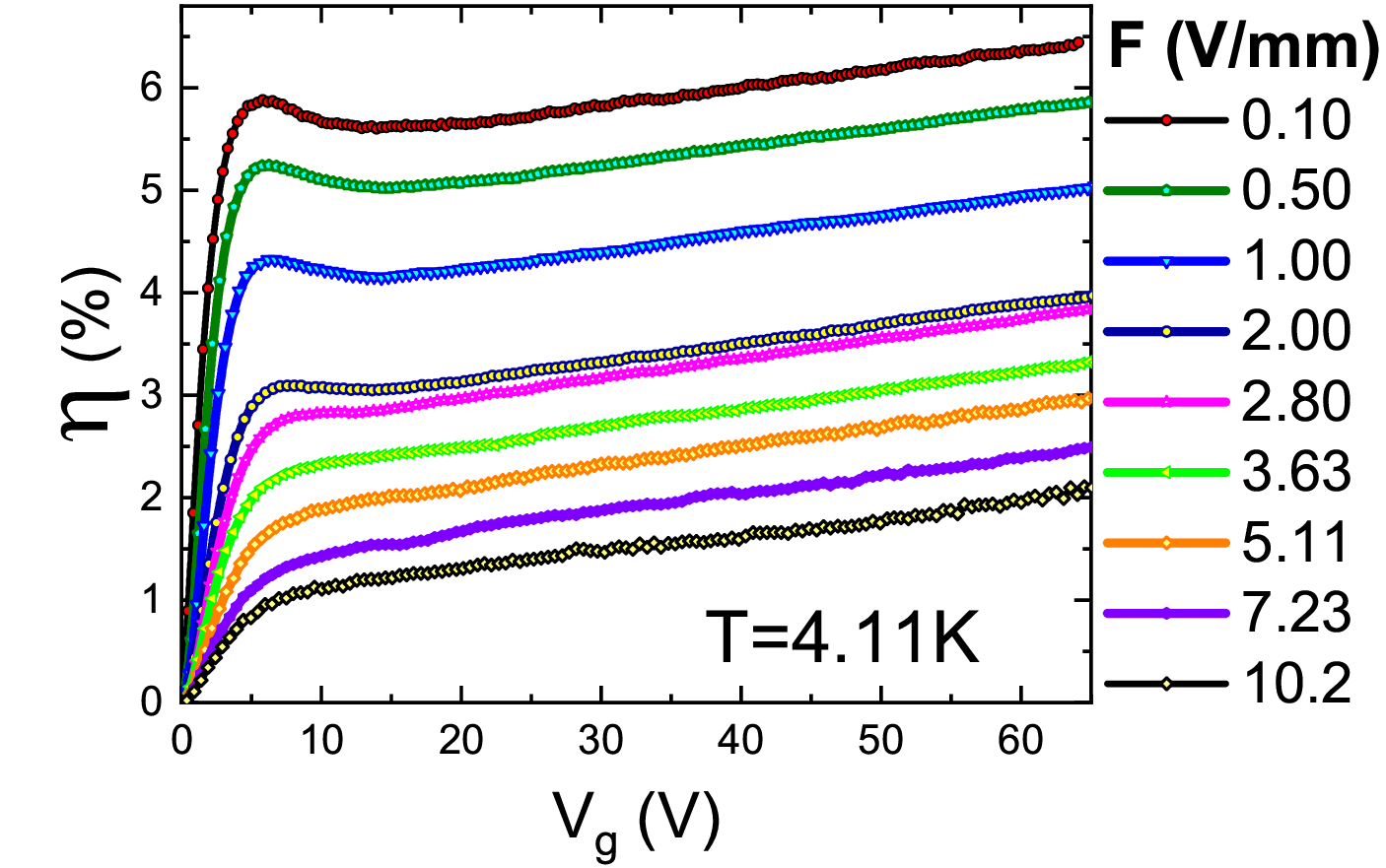}%
\caption{The same as in Fig. 3 except that the G(V$_{\text{g}}$) curves were
taken under various electric fields.}%
\end{figure}

Evidently, the two sets shown in Fig.~3 and Fig.~4 bear strong resemblance to
one another. The conditions under which these data were collected however, are
quite different; the set in Fig.~3 was taken at equilibrium while that in
Fig.~4 was taken as a NESS. The similarity in terms of how the G(V$_{\text{g}%
}$) curves evolve with either temperature or field may tempt one to assign a
T$_{\text{eff}}$ for a given G(F). At the same time, an effective temperature
T$_{\text{eff}}$(F) may be independently inferred from the data sets in Fig.~2
based on G$_{\text{0}}$. These two protocols for assigning a T$_{\text{eff}}%
$(F) use the \textit{same} set of measurements while focusing on different
parts of the G(V$_{\text{g}}$,F) data. They yield, however, quite different
values for T$_{\text{eff}}$ as is shown in Fig.~5 and Fig.~6 below.

Fig.~5 compares a set of $\eta^{\prime}$s measured at \textit{equilibrium}
temperatures with the set measured under NESS conditions produced by applying
different fields. The latter is plotted versus T$_{\text{eff}}$ calibrated
against the data in Fig.~2b on the basis of G$_{\text{0}}$ values. For brevity
sake, we shall refer to this set of T$_{\text{eff}}$(F) as that of thermometer
\#1. T$_{\text{eff}}$(F) for the thermometer \#2 is then obtained by finding
the temperature on the $\eta$(T) curve that fulfills: $\eta$(T$_{\text{eff}}%
$)=$\eta$(T) as illustrated in Fig.~5 for a specific $\eta$ value (red arrow).
This allows a direct comparison between the T$_{\text{eff}}$(F) of the two
thermometers depicted in Fig.~6.

It is not surprising that thermometer \#1 agrees with thermometer \#2
\textit{only} when F is vanishingly small. It is however remarkable that the
largest relative discrepancy between the thermometers is where the system is
just off the linear response regime in terms of G$_{\text{0}}$(F). This
observation demonstrates that T$_{\text{eff}}$ may be an ill-defined concept
in a non-equilibrium steady-state, even under a fairly weak drive. Such
hot-electron effects \cite{19} are commonly manifested in the transport
studies of semiconductors driven by non-Ohmic fields. The system conductance
under these conditions may also be influenced by electron-electron
correlations. %

\begin{figure}[ptb]%
\centering
\includegraphics[
height=2.693in,
width=3.4411in
]%
{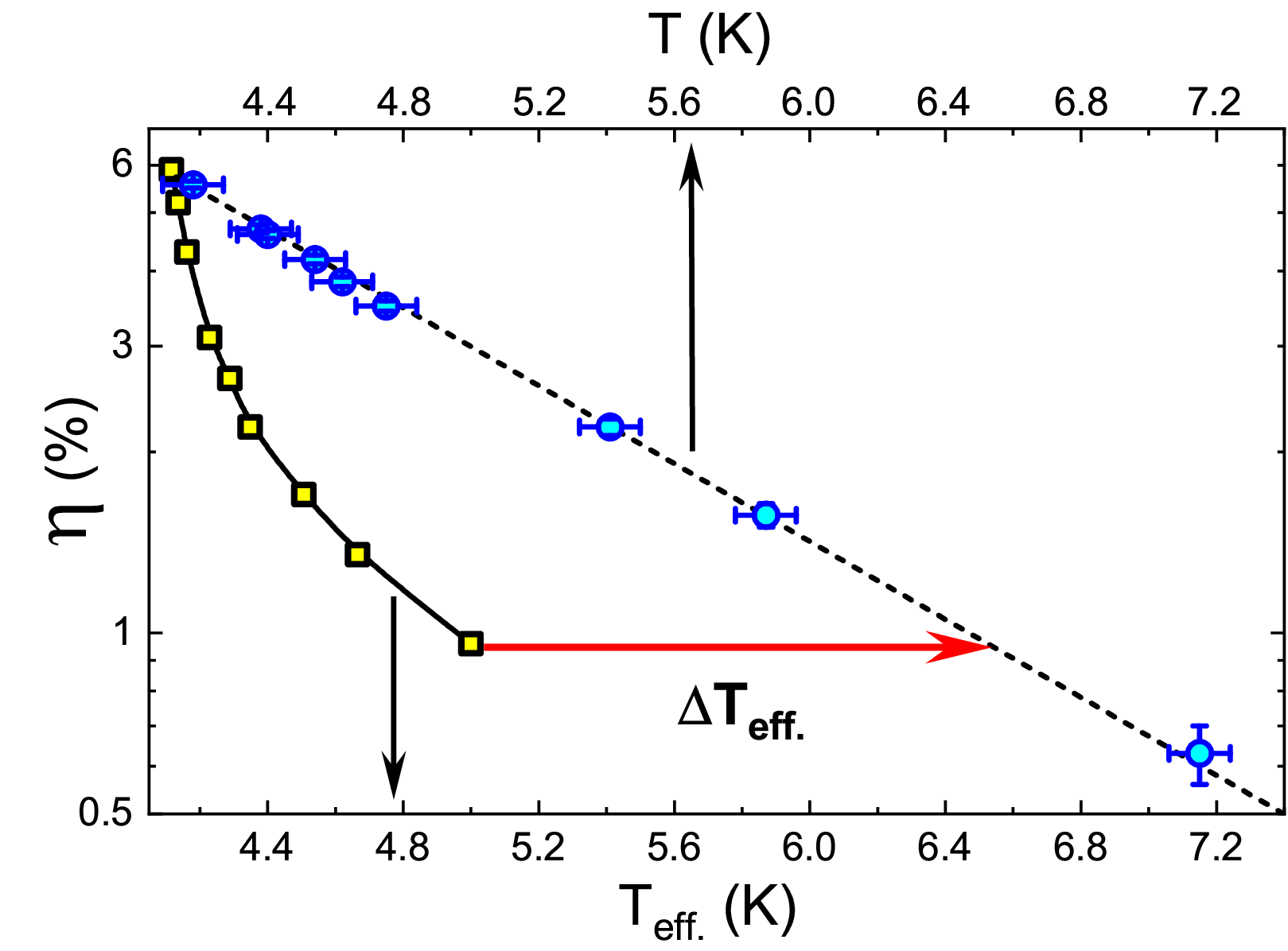}%
\caption{Contrasting the effective temperatures yielded by the two
thermometers under applied fields. Data for T$_{\text{eff}}$ for thermometer
\#1 (squares, and line as guide to the eye). Each $\eta$ taken from Fig. 4 is
assigned a T$_{\text{eff}}$ by using the data in Fig. 2 for G(T) and G(F) for
the sample. Mapping T$_{\text{eff}}$ for thermometer \#2 is accomplished by
matching the value of the measured $\eta$ with the equilibrium values for
$\eta$(T) taken from Fig. 3. The dashed line is a fit to: $\eta\propto\exp
$[-T/T*] with T*=1.34.}%
\end{figure}
%

\begin{figure}[ptb]%
\centering
\includegraphics[
height=2.5028in,
width=3.4411in
]%
{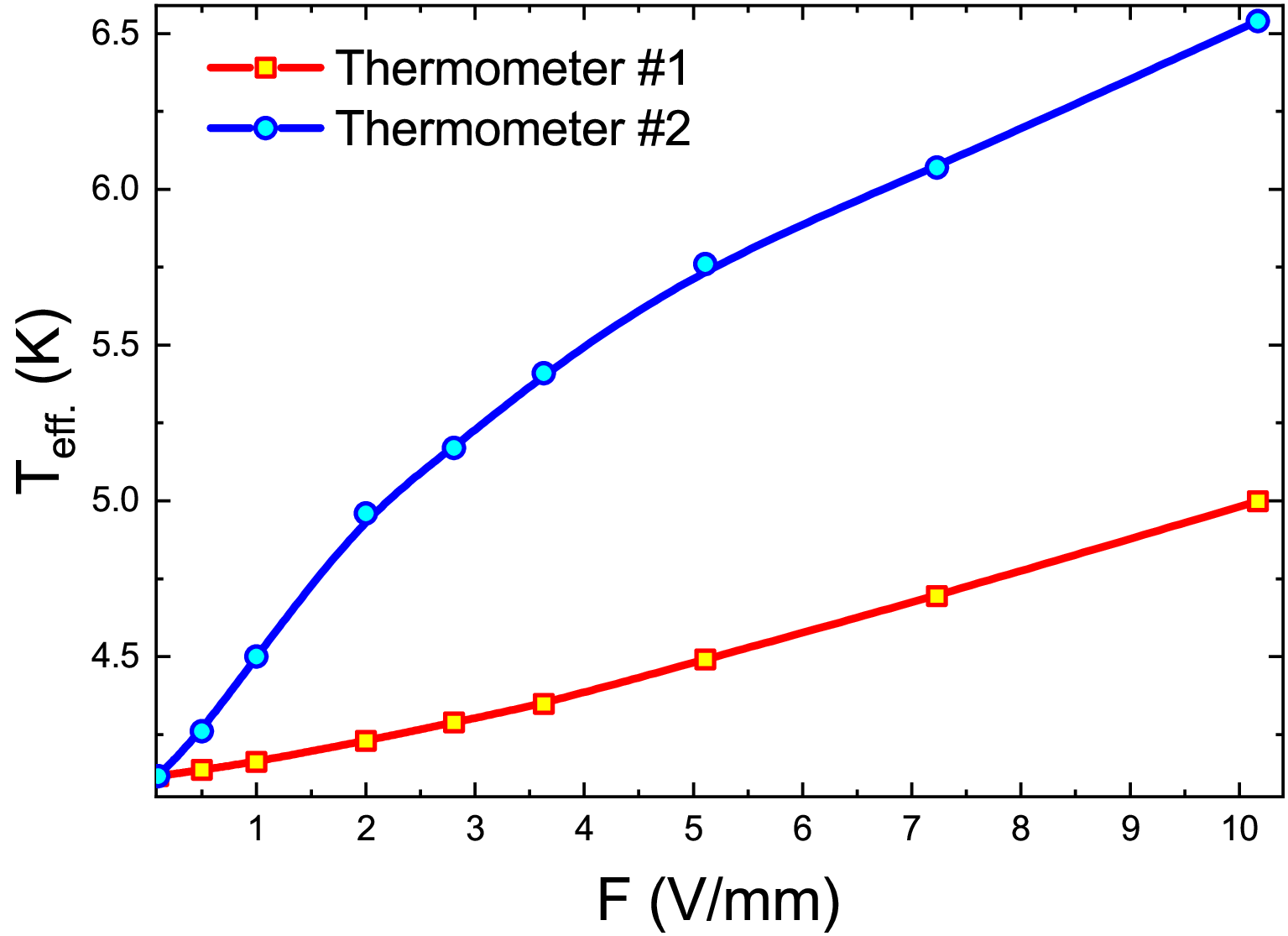}%
\caption{A direct comparison of the effective temperatures for thermometers
\#1 and \#2 based on data from Fig. 5 (see text).}%
\end{figure}

The difference between the two thermometers is probably even larger than
conveyed by Fig.~6. Note that T$_{\text{eff}}$ for thermometer \#1 is
determined here solely on the basis of the measured $\Delta$G(F). However,
$\Delta$G(F) is composed of \textit{two} contributions. The first is due to
field-assisted tunneling \cite{20,21,22,23}, the second results from
Joule-heating \cite{24,25,26}. The T$_{\text{eff}}$ used in Fig.~6 is
therefore larger than the value based on assuming just heating. This
complication, in different forms, is inherent to resistance-based thermometry
commonly used in experiments on disordered insulators.

Moreover, when the system is driven into a NESS by exposure to IR illumination
the discrepancy between the two T$_{\text{eff}}$'s turns out to be
significantly larger than under non-Ohmic fields. Figure 7a shows the raw data
for the equilibrium G(V$_{\text{g}}$) compared with a trace taken under weak
IR illumination.%

\begin{figure}[ptb]%
\centering
\includegraphics[
height=3.4091in,
width=3.4022in
]%
{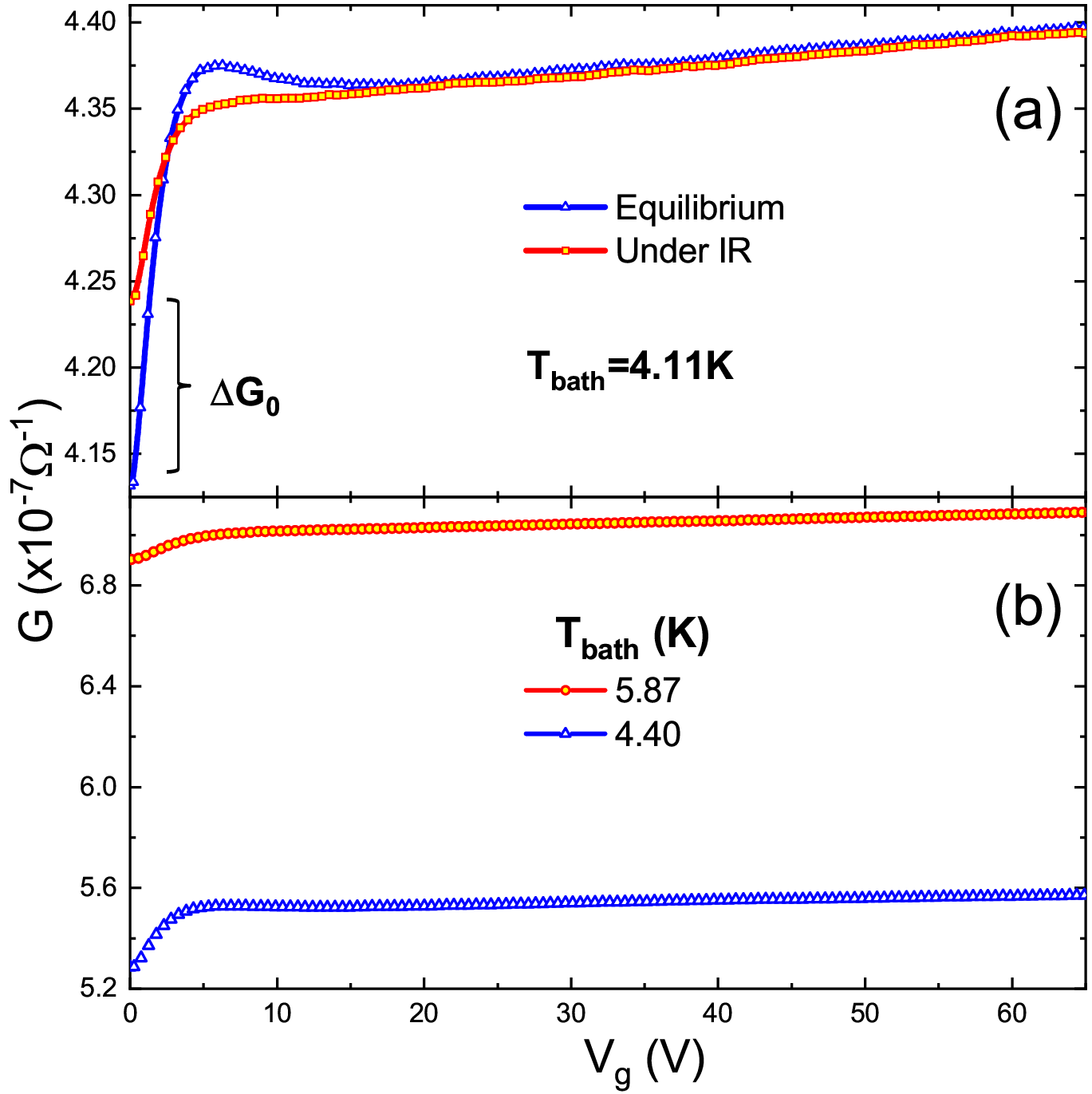}%
\caption{Conductance versus gate-voltage G(V$_{\text{g}}$) for the
In$_{\text{x}}$O sample under two different conditions. (a): Comparing
G(V$_{\text{g}}$) measured at equilibrium with a trace measured under IR
illumination. The IR source (operating at 0.2$\mu$A while biased at 1.003V)
has been on for 75 minutes before the trace was taken to ensure steady-state
conditions. (b): G(V$_{\text{g}}$) traces taken from the (equilibrium) data in
Fig. 3 as a pair that exhibit values of $\eta$ similar to these of the traces
in (a).}%
\end{figure}
Note first that the zero-bias conductance of the sample has slightly increased
in the NESS of the sample by $\Delta$G$_{\text{0}}$, a relative change of
$\approx$2.5\%. This is equivalent (using the calibration graph in Fig.~2) to
$\Delta$T$\approx$24mK while (using the data in Fig.~5) the change in $\eta$
is equivalent to $\Delta$T$_{\text{eff}}\approx$1.1K.

Secondly, To get a comparable shift of $\Delta$T$_{\text{eff}}$ requires an
applied non-Ohmic field F$\approx$3.7V/mm that, in turn, entails maintaining
input power into the sample of the order of 10$^{\text{-5}}$W. The
\textit{total} power used by the light-emitting diode in Fig.~7a is $\approx
$10$^{\text{-7}}$W \cite{27}. Obviously much less power is required by IR
illumination to reduce $\eta$ than by applying low frequency fields.

Another intriguing feature of the NESS maintained by IR illumination may be
appreciated by comparing the data in Fig.~7a with Fig.~7b. The latter
illustrates characteristic G(V$_{\text{g}}$) traces of the electron glass
measured in \textit{equilibrium}. As the bath temperature is raised, or when a
larger non-Ohmic F is applied, the \textit{entire} G(V$_{\text{g}}$) trace is
shifted to higher values. By contrast, only G$_{\text{0}}$ is enhanced when
the sample is exposed to IR while for V$_{\text{g}}\geq$2V, G(V$_{\text{g}}$)
falls below the equilibrium value. This feature was consistently observed on
three other samples using IR power $\lesssim$10$^{\text{-6}}$W.

To all appearances, the difference between the two NESS protocols depends on
the nature of the drive; non-Ohmic field versus IR-excitation.

The natural question is what makes IR illumination so effective in reducing
$\eta$ while barely affecting the system conductance. Additionally, why the
NESS created by non-Ohmic fields requires much larger power to achieve a
similar reduction in $\eta$?

The answer to both questions is related to the way the invested power of a
protocol modifies the phonon distribution D($\varepsilon$) of the system. In
equilibrium, D($\varepsilon$) is given by the Bose-Einstein distribution that,
at cryogenic temperatures, means an abundance of low-frequency phonons and
exponentially rare high-frequency ones.

Applying non-Ohmic fields F mostly adds low-frequency modes to the spectrum.
Down-going transitions associated with hopping under such fields produce
athermal phonons in the spectrum but with energies up to $\approx$eFL, where
L$_{\text{C}}$=L$_{\text{C}}$(F,T) is the percolation-radius
\cite{28,29,24,25}. Typical values for L$_{\text{C}}\approx$10$^{\text{-7}}$m
at T$\approx$4K, and fields in the range used in this work, the energy
accumulated from the field will appear as excess of phonons at few degrees
above the bath temperature.

When the system is exposed to the IR source the modification to D($\varepsilon
$) is more profound. The IR initiates a cascade process; electrons are excited
to high-energy, then relax by phonons emission \cite{31}. The NESS that
sets-in, sustains vibrations at phonon-energies of few hundred degrees
\cite{32}. Then, it is the high-energy part of D($\varepsilon$) that is
significantly boosted while the relative change in the low-energy phonon-bank
is small. The greater efficiency of the IR protocol relative to that of
non-Ohmic fields in reducing $\eta$ is a consequence of the excess high-energy
modes it creates. These modes allow transitions that, in equilibrium, would
occur only at temperatures of few hundred degrees where no memory-dip has ever
been observed in any of the half dozen \cite{18} electron-glasses measured todate.

The two thermometers also differ in the typical energy controlling their
response. For thermometer \#1 (based on the measured conductance), this is the
optimal hopping-energy \cite{33,34} E$_{\text{opt}}$=(k$_{\text{B}}%
$T)$^{\text{2/3}}$($\partial$n$/\partial\mu$\textperiodcentered
d\textperiodcentered$\xi^{\text{2}}$)$^{\text{-1/3}}$. Here, $\partial
$n$/\partial\mu$ $\approx$10$^{\text{32}}$erg$^{\text{-1}}$cm$^{\text{-3}}$ is
the density of states, d is the film thickness, and $\xi\approx$5nm is the
localization length (estimated from G(T) in Fig.~2b). With these values
E$_{\text{opt}}$ is of order $\approx$1meV at $\approx$5K. The characteristic
energy for thermometer \#2 is the disorder energy W$\gtrsim$0.4eV. This is
estimated on the basis of the condition that W is large enough to
Anderson-localize the system and to slow its relaxation such that the
memory-dip is resolved in the field-effect scans \cite{18}.

Hopping conductivity takes place in a current-carrying network (CNN) utilizing
regions of the relatively weak disorder. The CNN encompasses regions of the
highest disorder in the sample \cite{28,29,30}. The large disorder in these
regions is responsible for the slow relaxation of the electron glass
\cite{18}. The two regions \cite{35} communicate mainly via low-energy phonons
\cite{18}, and electron-electron interaction. The range over which interaction
is effective however may depend on available resonances and therefore on
quantum coherence in the medium \cite{36}.

In sum, we have demonstrated that driving an Anderson insulator even slightly
out of its linear response entails a sharp thermometers-conflict suggesting
that thermometry in this regime \cite{37} should be treated with some doubt.
The frequency of the drive turns out to be a pivotal parameter in the
thermometers discrepancy. Explicitly, the difference between the thermometers
is significantly larger when the drive initiates high-frequencies energy
quanta in the medium as occurs in the IR protocol. A specific reason for this
was offered based on the quantum nature of the phonon energy-distribution.
Phonons are an essential ingredient in establishing steady-state conditions in
these NESS experiments. Our results show that they also play a nontrivial role
in nonequilibrium phenomena.

\begin{acknowledgments}
The author thanks Preeti Bhandari, Vikas Malik, and Moshe Schechter for
illuminating discussions on their simulation work.
\end{acknowledgments}


\begin{thebibliography}{99}
\bibitem {27}Note that only part of this power is actually invested in the
sample due to imperfect efficiency, optical absorption, beam-geometry, etc.
\end{thebibliography}
\end{document}